\documentclass[twocolumn,american]{revtex4-2}
\usepackage[T1]{fontenc}
\usepackage[utf8]{luainputenc}
\usepackage{geometry}
\usepackage{amsmath}
\usepackage{amssymb}
\usepackage{babel}
\begin{document}
\title{Failure of the Quench Action Formalism for Mott Insulator Initial
States}
\author{Garry Goldstein}
\address{garrygoldsteinwinnipeg@gmail.com}
\begin{abstract}
The quench action formalism relies on the assumption that the overlap
between a generic initial state $\left|\Psi_{0}\right\rangle $ and
an eigenstate of an integrable model - defined through the rapidities
$\left|k_{1},...k_{N}\right\rangle $ - can be written as: 
\begin{equation}
\left\langle k_{1},...k_{N}\mid\Psi_{0}\right\rangle =\exp\left(-S_{\Psi_{0}}\left(\rho\left(k\right)\right)\right),\label{eq:Exponential}
\end{equation}
where $\rho\left(k\right)$ is the quasiparticle density of the state
$\left|k_{1},...k_{N}\right\rangle $ and $S_{\Psi_{0}}$ is some
smooth function of $\rho\left(k\right)$ that depends on $\Psi_{0}$.
In particular the quench action formalism assumes the overlap depends
smoothly on the quasiparticle density $\rho\left(k\right)$. In this
work, by explicit counter example, we show that this is not the case.
We consider the quench between a Mott insulator and a Lieb Liniger
gas. We show that the overlap between the ground state of the Mott
insulator and arbitrary eigenstates of the Lieb Liniger gas has a
highly singular behavior and no expression like Eq. (\ref{eq:Exponential})
applies. We do so within the Tonks Girardeau limit of the Lieb Liniger
gas and to leading order in the $1/c$ expansion for the overlap (with
$c$ being the coupling constant of the Lieb Liniger gas). In Appendix
\ref{sec:XXZ-Model-for} we show similar results for overlaps in the
XXZ model with crystal states.
\end{abstract}
\maketitle

\section{Introduction}\label{sec:Introduction}

The question of how and whether an isolated quantum system relaxes
to which equilibrium state is of great interest to modern physicist.
On the forefront of this area of research are studies of Bethe integrable
systems \citep{Bethe_1931,Zvyagin_2005,Korepin_1993,Samaj_2013,Franchini_2017}
in 1D. In Bethe integrable systems the exact eigenstates are known
due to the existence of an infinite number of commuting conserved
quantities $Q_{i}$ \citep{Samaj_2013,Zvyagin_2005,Franchini_2017}.
In particular the eigenstates can be characterized in terms of rapidities
$\left|k_{1},...k_{N}\right\rangle $. Furthermore there has been
tremendous experimental effort to understand the dynamics and thermodynamics
of these integrable systems \citep{Weiss_2004,Amerogen_2008,Dahr_2025,Ganahal_2012,Gring_2012,Haller_2009,Kinoshita_2006,Langen_2015,Paredez_2004,Rosenberg_2024,Trotzky_2012,Meinert_2015}.
One of the key questions asked is the fate of a quenched system. A
quench is when a system is initially prepared in an eigenstate $\left|\Psi_{0}\right\rangle $
of a Hamiltonian $H_{0}$ and then instantaneously - on the time scale
of the experiment - a new Hamiltonian $H_{f}$ is turned on and the
system evolves as: 
\begin{equation}
\left|\Psi\left(t\right)\right\rangle =\exp\left(-iH_{f}t\right)\left|\Psi_{0}\right\rangle .\label{eq:Evolves}
\end{equation}
Here $t$ is the time of evolution. One of the key questions involved
is what happens if $H_{f}$ is an integrable Hamiltonian. By now it
is strongly theoretically believed that the final state of the system
is given by a Generalized Gibbs Ensemble (GGE) \citep{Rigol_2008,Goldstein_2014(2),Ilievski_2015,Pozsgay_2014},
with the GGE being observed experimentally in Ref. \citep{Langen_2015}.
One of the key theoretical underpinnings of the GGE formalism and
the way in which an integrable system equilibrates to the GGE is the
quench action formalism \citep{Caux_2013,Caux_2016,Brockmann_2014,Brockmann_2014(2),Brockmann_2014(3),Ilievski_2015,Luca_2015,Mastyan_2015,Nardis_2014,Wouters_2014,Wouters_2015}.
In the quench action formalism it is assumed that the overlap between
the initial state and an eigenstate of the integrable final Hamiltonian
is given by:
\begin{equation}
\left\langle k_{1},...k_{N}\mid\Psi_{0}\right\rangle =\exp\left(-S_{\Psi_{0}}\left(\rho\left(k\right)\right)\right)\label{eq:Exponential-1}
\end{equation}
Where $\rho\left(k\right)$ is a the quasiparticle density of the
state $\left|k_{1},...k_{N}\right\rangle $ and $S_{\Psi_{0}}$ is
some smooth function of $\rho\left(k\right)$ which is dependent on
$\left|\Psi_{0}\right\rangle $. The final state at long times $\left|\Psi\left(t\rightarrow\infty\right)\right\rangle $
is then shown to be essentially an eigenstates of $H_{f}$ and the
quasiparticle density of the final eigenstate of $H_{f}$ is determined
by maximizing the Yang-Yang action for the system given by: 
\begin{equation}
Z=\int\mathcal{D}\left(\rho\left(k\right)\right)\exp\left(-2Re\left[S_{\Psi_{0}}\left(\rho\left(k\right)\right)\right]+S_{YY}\left(\rho\left(k\right)\right)\right)\label{eq:Partition}
\end{equation}
with respect to $\rho\left(k\right)$, subject to the Bethe ansatz
equations for $\rho\left(k\right)$. Here $S_{YY}\left(\rho\left(k\right)\right)$
is the Yang-Yang entropy of the system \citep{Yang_1969,Zvyagin_2005,Samaj_2013,Korepin_1993,Franchini_2017}.
More precisely the long term behavior of the integrable system is
controlled by the density $\rho^{*}\left(k\right)$ s.t.

\begin{equation}
\frac{\delta}{\delta\rho\left(k\right)}\left[-2Re\left[S_{\Psi_{0}}\left(\rho^{*}\left(k\right)\right)\right]+S_{YY}\left(\rho^{*}\left(k\right)\right)\right]=0\label{eq:Saddle_equation}
\end{equation}
Furthermore equilibration can be understood by studying small fluctuations
about the saddle point $\rho^{*}\left(k\right)$ \citet{Caux_2013,Caux_2016}.
Models with multiple species of quasiparticles (such as bound states)
can be studied by replacing $\rho\left(k\right)\rightarrow\rho_{n}\left(k\right)$
in Eqs. (\ref{eq:Exponential-1}), (\ref{eq:Partition}) and (\ref{eq:Saddle_equation}),
where $n$ labels the different quasiparticle species. 

The main result of this work is that this approach is incorrect, the
quench action formalism is wrong. The most pertinent mistake of the
quench action formalism is that the quench action formalism assumes
that if we take an initial state $\left|\Psi_{0}\right\rangle $ then
the overlap with the state $\left|k_{1},k_{2},...k_{N}\right\rangle $
(with $N/L$ fixed, where $L$ is the length of the system) depends
smoothly on the density of states $\rho\left(k\right)$. Thats is
if we take two eigenstates of $H_{f}$, $\left|\Phi_{1}\right\rangle $
and $\left|\Phi_{2}\right\rangle $, with the same $\rho\left(k\right)$,
the overlap does not depend on which states we are considering: 
\begin{equation}
\left\langle \Phi_{1}\mid\Psi_{0}\right\rangle =\left\langle \Phi_{2}\mid\Psi_{0}\right\rangle +O\left(\frac{1}{N}\right)\label{eq:Equals}
\end{equation}
We will show this through a counter example to be explicitly false
and that overlaps have significantly more structure then in given
in Eq. (\ref{eq:Exponential-1}) above. We first would like to note
that there is already a counter example to Eq. (\ref{eq:Equals})
in the literature \citep{Brockmann_2014,Brockmann_2014(2),Brockmann_2014(3),Caux_2016,Nardis_2014,Wouters_2014,Wouters_2015}.
For this counter example we can take an integrable quench \citep{Pozsgay_2020,Piroli_2017}
- where for the conserved charges (for say the Lieb Liniger gas) we
have that: 
\begin{equation}
Q_{2p+1}\left|k_{1},k_{2},...k_{N}\right\rangle =\sum_{i}k_{i}^{2p+1}\left|k_{1},k_{2},...k_{N}\right\rangle \label{eq:zero}
\end{equation}
 and the initial state $\left|\Psi_{0}\right\rangle $ satisfies:
\begin{equation}
Q_{2p+1}\left|\Psi_{0}\right\rangle =0,\:\forall p\label{eq:Conserved_zero}
\end{equation}
Then: 
\begin{align}
 & \left\langle k_{1},k_{2},...k_{N}\mid\Psi_{0}\right\rangle =0,\nonumber \\
 & unless:k_{1}=-k_{2},\:k_{3}=-k_{4},\:k_{5}=-k_{6},...\label{eq:pair_structure}
\end{align}
(there is paring of the rapidities). For such states there is an ad
hoc correction to the quench action - where the Yang-Yang entropy
is divided by two:
\begin{equation}
S_{YY}\left(\rho\left(k\right)\right)\rightarrow\frac{1}{2}S_{YY}\left(\rho\left(k\right)\right)\label{eq:Entropy}
\end{equation}
in Eqs. (\ref{eq:Partition}) and (\ref{eq:Saddle_equation}). In
this work we show that the problem is much worse that there are overlaps
with significantly more structure then just pairing where there is
no obvious modification to the Yang-Yang entropy that will work. We
will do so by an explicit example of the quench of the Mott insulator
into the Tonks Girardeau gas \citep{Girardeau_1960,Tonks_1936} and
to leading order in $1/c$ for overlaps of the Lieb Liniger gas \citep{Lieb_1963,Lieb_1963(2)}
(here $c$ is the coupling constant of the Lieb Liniger gas). In Appendix
\ref{sec:XXZ-Model-for} we will study similar states (crystal states)
for the XXZ model, in the limit a large anisotropy $\left|\Delta\right|\gg1$,
and find they also have a very rich overlap structure where Eq. (\ref{eq:Equals})
does not apply.

\section{Explicit Example: Quench between a Mott Insulator and the Tonks
limit of the Lieb Liniger gas}\label{sec:Explicit-Example}

The Lieb Liniger Hamiltonian is given by: 
\begin{equation}
H=\int_{0}^{L}dx\left[\partial_{x}b^{\dagger}\left(x\right)\partial_{x}b\left(x\right)+cb^{\dagger}\left(x\right)b^{\dagger}\left(x\right)b\left(x\right)b\left(x\right)\right]\label{eq:Hamiltonian_lieb_liniger}
\end{equation}
Here $b^{\dagger}\left(x\right)$, $b\left(x\right)$ are bosonic
creation and annihilation operators at the point $x$. We will assume
periodic boundary conditions. We consider the initial state \citep{Goldstein_2014}:
\begin{equation}
\left|\Psi_{0}\right\rangle =\prod_{j=0}^{N-1}\int_{-\infty}^{\infty}dx\varphi\left(x-jl\right)b^{\dagger}\left(x\right)\left|0\right\rangle \label{eq:Initial_state-3}
\end{equation}
with $Nl=L$. Where we will assume that the width $W$ of $\varphi\left(x\right)$
is much small the $l$, $W\ll l$ so that 
\begin{equation}
\int_{-\infty}^{\infty}dx\varphi^{*}\left(x-jl\right)\varphi\left(x-kl\right)\cong\delta_{jk}\label{eq:Delta}
\end{equation}
Where if we further assume that: 
\begin{equation}
\varphi\left(x\right)=\frac{\exp\left(-x^{2}/\sigma\right)}{\left(\pi\sigma/2\right)^{1/4}}\label{eq:Gaussian-1}
\end{equation}
with $\sqrt{\sigma}\ll l$ we get the Mott insulator initial state
quench \citep{Goldstein_2014}. We will show that the state in Eq.
(\ref{eq:Initial_state-3}) has much richer structure of overlaps
with the eigenstates of the Lieb Liniger gas then is given by Eq.
(\ref{eq:Equals}) above. We will not use the assumption given in
Eq. (\ref{eq:Gaussian-1}) above. Then the un-normalized eigenstates
of the Lieb Liniger gas are given by:
\begin{align}
 & \left|k_{1},..k_{N}\right\rangle \nonumber \\
 & =\int_{0<x_{1}<...x_{N}<L}\psi\left(k_{1},...k_{N};x_{1},...x_{N}\right)\prod_{j}b^{\dagger}\left(x_{j}\right)\left|0\right\rangle \label{eq:Wavefunction_Bethe}
\end{align}
with:
\begin{align}
\psi\left(k_{1},...k_{N};x_{1},...x_{N}\right) & =\sum_{P\in\pi_{N}}\left(-1\right)^{P}\exp\left[\sum_{j}ik_{Pj}x_{j}\right]\nonumber \\
 & \times\prod_{m<n}\left(k_{Pm}-k_{Pn}+ic\right)\label{eq:Definition}
\end{align}
Where $\pi_{N}$ is the permutation group on $N$ elements. To find
the overlap between the wavefunction in Eq. (\ref{eq:Initial_state-3})
and (\ref{eq:Wavefunction_Bethe}) we first need to integrate:
\begin{equation}
\int\varphi\left(x-jl\right)\exp\left(-ik_{Pj}\left(x\right)\right)=\exp\left(-ik_{Pj}jl\right)\tilde{\varphi}\left(k_{Pj}\right)\label{eq:Fourier}
\end{equation}
where $\tilde{\varphi}\left(k\right)=\int dxe^{-ikx}\varphi\left(x\right)$
is the Fourier transform of $\varphi\left(x\right)$. Now doing the
Wick contractions we have that: 
\begin{align}
\left\langle k_{1},....k_{N}\mid\Psi_{0}\right\rangle  & =\prod_{j}\tilde{\varphi}\left(k_{j}\right)\sum_{P\in\pi_{N}}\left(-1\right)^{P}\exp\left[-i\sum_{j}jlk_{Pj}\right]\nonumber \\
 & \times\prod_{m<n}\left(k_{Pm}-k_{Pn}-ic\right)\label{eq:Overlap_2}
\end{align}
Lets take $c\rightarrow\infty$ and study: 
\begin{align}
 & \frac{1}{\left(-ic\right)^{N\left(N-1\right)/2}}\left\langle k_{1},....k_{N}\mid\Psi_{0}\right\rangle \nonumber \\
 & =\prod_{j}\tilde{\varphi}\left(k_{j}\right)\sum_{P\in\pi_{N}}\left(-1\right)^{P}\exp\left(-i\sum_{j}jlk_{Pj}\right)\nonumber \\
 & =\prod_{j}\tilde{\varphi}\left(k_{j}\right)\det\left(\exp\left(-ilk_{i}\right)^{j}\right)\nonumber \\
 & =\prod_{j}\tilde{\varphi}\left(k_{j}\right)\prod_{j>i}\left(\exp\left(-ik_{j}l\right)-\exp\left(-ik_{i}l\right)\right)\label{eq:Vandermonde}
\end{align}
(we recognize the Vandermonde determinant). Now we study: 
\begin{align}
 & \frac{1}{c^{N(N-1)}}\left|\left\langle k_{1},....k_{N}\mid\Psi_{0}\right\rangle \right|^{2}\nonumber \\
 & =\prod_{j}\left|\tilde{\varphi}\left(-k_{j}\right)\right|^{2}\prod_{j>i}4\sin^{2}\left[\frac{k_{j}-k_{i}}{2}l\right]\label{eq:Product_sine-1}
\end{align}
Now we have that $lN=L$ and $k_{j}=\frac{2\pi I_{j}}{L}$ (with $I_{j}\in\mathbb{Z}$
we will assume $N$ is odd) so we have that: 
\begin{align}
 & \frac{1}{c^{N\left(N-1\right)}}\left|\left\langle k_{1},....k_{N}\mid\Psi_{0}\right\rangle \right|^{2}\nonumber \\
 & =\prod_{j}\left|\tilde{\varphi}\left(k_{j}\right)\right|^{2}\prod_{j>i}4\sin^{2}\left[\pi\frac{I_{j}-I_{i}}{N}\right]\label{eq:Overlap_Tonks}
\end{align}
We notice that this function is not a smooth function of $\rho\left(k\right)$.
Indeed we must have that every $I=0,1,2,....N-1$ (mod $N)$ is occupied
exactly once or the product in Eq. (\ref{eq:Overlap_Tonks}) is zero.
As such:
\begin{align}
 & \frac{1}{c^{N\left(N-1\right)}}\left|\left\langle k_{1},....k_{N}\mid\Psi_{0}\right\rangle \right|^{2}\nonumber \\
 & =\prod_{j}\left|\tilde{\varphi}\left(k_{j}\right)\right|^{2}\prod_{N\geq j>i\geq1}4\sin^{2}\left[\pi\frac{j-i}{N}\right]\label{eq:Overlap_Tonks_Simplified}
\end{align}
 for $\left\{ k_{1},....k_{N}\right\} =\left\{ p\frac{2\pi}{L}+m_{p}\frac{2\pi N}{L}\right\} ,\:with\:p=0,1,2,...N-1,\:m_{p}\in\mathbb{Z}$.
In other words the overlap has structure where only:
\begin{align}
 & \left\{ k_{1},....k_{N}\right\} =\left\{ p\frac{2\pi}{L}+m_{p}\frac{2\pi N}{L}\right\} ,\nonumber \\
 & \:with\:p=0,1,2,...N-1,\:m_{p}\in\mathbb{Z}\label{eq:Conditions}
\end{align}
have non-zero overlap. This is a much richer structure then in Eq.
(\ref{eq:Equals}). As such the quench action fails as Eq. (\ref{eq:Equals})
fails.

\section{ Mott Insulator Quench: Leading Order in $1/c$ expansion}\label{sec:Leading-Order-in}

We will again consider the Mott insulator quench but now to leading
order in $1/c$ expansion for the overlap. We will consider the initial
state given in Eq. (\ref{eq:Initial_state-3}) without the assumption
given in Eq. (\ref{eq:Gaussian-1}). Then the overlap of the initial
state with an eigenstate of the form in Eq. (\ref{eq:Wavefunction_Bethe})
is given by Eq. (\ref{eq:Overlap_2}). Now lets write:
\begin{equation}
k_{Pm}-k_{Pn}-ic\cong-ic\exp\left(i\left(k_{Pm}-k_{Pn}\right)/c\right)+O\left(\frac{1}{W^{2}c^{2}}\right)\label{eq:Approximate}
\end{equation}
This means that:
\begin{align}
 & \prod_{m<n}\left(k_{Pm}-k_{Pn}-ic\right)\nonumber \\
 & \cong\left(-ic\right)^{N\left(N-1\right)/2}\prod_{m}\exp\left(i\left[N-2m\right]k_{Pm}/c\right)\label{eq:Simpler_product}
\end{align}
So that:
\begin{align}
 & \frac{1}{\left(-ic\right)^{N\left(N-1\right)/2}}\left\langle k_{1},....k_{N}\mid\Psi_{0}\right\rangle \nonumber \\
 & =\exp\left(i\frac{N}{c}\sum_{i}k_{i}\right)\prod_{j}\tilde{\varphi}\left(k_{j}\right)\times\nonumber \\
 & \times\sum_{P\in\pi_{N}}\left(-1\right)^{P}\exp\left(-i\sum_{j}j\left[l+\frac{2}{c}\right]k_{Pj}\right)\nonumber \\
 & =\exp\left(i\frac{N}{c}\sum_{i}k_{i}\right)\prod_{j}\tilde{\varphi}\left(k_{j}\right)\det\left(\exp\left(-i\left[l+\frac{2}{c}\right]k_{i}\right)^{j}\right)\nonumber \\
 & =\exp\left(i\frac{N}{c}\sum_{i}k_{i}\right)\prod_{j}\tilde{\varphi}\left(k_{j}\right)\times\nonumber \\
 & \times\prod_{j>i}\left(\exp\left(-ik_{j}\left[l+\frac{2}{c}\right]\right)-\exp\left(-ik_{i}\left[l+\frac{2}{c}\right]\right)\right)\label{eq:Vandermonde_1/c}
\end{align}
(we recognize the Vandermonde determinant). Now we study: 
\begin{align}
 & \frac{1}{c^{N(N-1)}}\left|\left\langle k_{1},....k_{N}\mid\Psi_{0}\right\rangle \right|^{2}\nonumber \\
 & =\prod_{j}\left|\tilde{\varphi}\left(k_{j}\right)\right|^{2}\prod_{j>i}4\sin^{2}\left[\frac{k_{j}-k_{i}}{2}\left[l+\frac{2}{c}\right]\right]\label{eq:Det_sine}
\end{align}
Now we have that the Bethe equations are given by: 
\begin{align}
 & k_{j}L=2\pi I_{j}+\left(N-1\right)\pi-2\sum_{m}\arctan\left(\frac{k_{j}-k_{m}}{c}\right)\nonumber \\
 & \Rightarrow k_{j}\left(L+\frac{2N}{c}\right)\cong2\pi I_{j}+\left(N-1\right)\pi+2\sum_{m}\frac{k_{m}}{c}\nonumber \\
 & \Rightarrow\left[k_{j}-k_{i}\right]\left(L+\frac{2N}{c}\right)=2\pi\left(I_{j}-I_{i}\right)\label{eq:Bethe_equations}
\end{align}
Here $I_{i}$ are the occupied quantum numbers (with $I_{j}\in\mathbb{Z}$).
This means that: 
\begin{align}
 & \frac{1}{c^{N(N-1)}}\left|\left\langle k_{1},....k_{N}\mid\Psi_{0}\right\rangle \right|^{2}\nonumber \\
 & =\prod_{j}\left|\tilde{\varphi}\left(k_{j}\right)\right|^{2}\prod_{j>i}4\sin^{2}\left[\frac{\pi\left(I_{j}-I_{i}\right)}{N\left(l+\frac{2}{c}\right)}\left[l+\frac{2}{c}\right]\right]\nonumber \\
 & =\prod_{j}\left|\tilde{\varphi}\left(k_{j}\right)\right|^{2}\prod_{j>i}4\sin^{2}\left[\frac{\pi\left(j-i\right)}{N}\right]+O\left(\frac{1}{W^{2}c^{2}}\right)\label{eq:Same_before_det_sine}
\end{align}
where again we must have that every $I=0,1,2,....N-1$ (mod $N)$
is occupied exactly once or the overlap is zero. So the overlap is
finite if:
\begin{equation}
k_{j}=k_{1}+\frac{2\pi\left(j-1\right)}{L+\frac{2N}{c}}+n_{j}\frac{2\pi N}{L+\frac{2N}{c}},\:n_{j}\in\mathbb{Z}\label{eq:k_j}
\end{equation}
and zero otherwise. Again this is a much richer structure then in
Eq. (\ref{eq:Equals}). As such the quench action fails as Eq. (\ref{eq:Equals})
fails.

\section{Conclusions}\label{sec:Conclusions}

In this work we have presented a counterexample to the quench action
formalism. We have presented an initial state - the bosonic Mott insulator
ground state in 1D - whose overlap with the Lieb Liniger eigenstates
has a highly singular form - where Eq. (\ref{eq:Exponential}) does
not apply in any form. In Appendix \ref{sec:XXZ-Model-for} we study
some initial states (crystal states) for quenches of the XXZ model,
in the limit of large anisotropy $\left|\Delta\right|\gg1$, and find
similar results where Eq. (\ref{eq:Equals}) fails. We would like
to note that this is a very exciting development - whereby overlaps
have significantly more structure then expectation values of local
operators - which are known to be smooth functions of $\rho\left(k\right)$
\citep{Korepin_1993}. The quench action, which previously was thought
to describe all quenches into integrable models, apparently greatly
simplified the situation - making all quenches appear somewhat bland;
quenches were always controlled by a version of the Yang-Yang equations
and small fluctuations about the saddle point of the Yang-Yang action.
Now we see that the situation, in reality, is much more rich and exciting
with many singular possibilities leading to potentially new formalisms
and new theoretical approaches as to how or if the system equilibrates
to a GGE. To support our claims further, in the future, it would be
of interest to look into nearly integrable quenches, ones with initial
states of the form: $\left|\Psi_{0}\right\rangle =$ 
\begin{equation}
\left(\frac{1}{\sqrt{L\left(1+\left|\alpha\right|^{2}\right)}}\int_{0}^{L}dx\left(1+\alpha\exp\left(ikx\right)\right)b^{\dagger}\left(x\right)\right)^{N}\left|0\right\rangle \label{eq:Inital_BEC}
\end{equation}
with $\left|\alpha\right|\ll1$ and obtain conditions similar to Eq.
(\ref{eq:pair_structure}) but with deviations of the order of $\alpha$
leading to more complex corrections to the Yang-Yang entropy then
in Eq. (\ref{eq:Entropy}) thereby presenting another counter-example
to the quench action formalism.

\textbf{Acknowledements:} The author would like to thank Pradip Kattel
and Natan Andrei for useful discussions.

\appendix

\section{ XXZ Model for $\left|\Delta\right|\gg1$: Crystal State Quenches}\label{sec:XXZ-Model-for}

In this appendix we will consider crystal state quenches in to the
XXZ model. We consider the XXZ model with $\left|\Delta\right|\gg1$
with periodic boundary conditions on a ring of length $L$. The Hamiltonian
of the XXZ model is given by: 
\begin{equation}
H_{XXZ}=J\sum_{j=1}^{L}\left(\sigma_{j}^{x}\sigma_{j+1}^{x}+\sigma_{j}^{y}\sigma_{j+1}^{y}+\Delta\sigma_{j}^{z}\sigma_{j+1}^{z}\right)\label{eq:XXZ_Hamiltonian}
\end{equation}
(with $\vec{\sigma}_{L+1}\equiv\vec{\sigma}_{1}$). Here $\sigma_{i}^{x/y/z}$
are the Pauli matrices at site $i$ and $J$ is a irrelevant coupling
constant. Now consider the initial state (crystal state): 
\begin{equation}
\left|\Psi_{0}\right\rangle =\prod_{j=1}^{N}\sigma_{jl}^{-}\left|\uparrow......\uparrow\right\rangle \label{eq:Initial_state-2}
\end{equation}
Where $l=\frac{L}{N}\in\mathbb{N}$. Then we have that the un-normalized
eigenstates of the XXZ Hamiltonian are given by \citep{Korepin_1993}:
$\left|\lambda_{1},....\lambda_{N}\right\rangle =$

\begin{equation}
\sum_{m_{1}<m_{2}<....m_{N}}\chi\left(\lambda_{1},....\lambda_{N};m_{1},....m_{N}\right)\sigma_{m_{1}}^{-}....\sigma_{m_{N}}^{-}\left|\uparrow......\uparrow\right\rangle \label{eq:Operator_wavefunction}
\end{equation}
where:
\begin{align}
 & \chi\left(\lambda_{1},....\lambda_{N};m_{1},....m_{N}\right)\nonumber \\
 & =\sum_{P\in\pi_{N}}\left(-1\right)^{P}A_{P}\exp\left(i\sum_{a=1}^{N}\lambda_{Pa}m_{a}\right)\label{eq:XXZ_wavefunction}
\end{align}
We will assume $\lambda_{a}\in\mathbb{R}$ that is neglect states
with string solutions. This will already be enough to show the failure
of Eq. (\ref{eq:Exponential-1}) for those states.
\begin{widetext}
Where:
\begin{align}
A_{P} & =\left(-1\right)^{N-1}\prod_{j<m}\left(\exp\left(i\left[\lambda_{Pm}+\lambda_{Pj}\right]\right)+1-2\Delta\exp\left(i\lambda_{Pj}\right)\right)\nonumber \\
 & =\left(-1\right)^{N-1}\left(-2\Delta\right)^{\left(N-1\right)N/2}\prod_{k}\exp\left(i\left(N-k\right)\lambda_{Pk}\right)\prod_{j<m}\left(1-\frac{\exp\left(i\lambda_{Pm}\right)}{2\Delta}-\frac{\exp\left(-i\lambda_{Pj}\right)}{2\Delta}\right)\nonumber \\
 & \cong\left(-1\right)^{N-1}\left(-2\Delta\right)^{\left(N-1\right)N/2}\prod_{k}\exp\left(i\left(N-k\right)\lambda_{Pk}\right)\prod_{j<m}\left(\exp\left[-\frac{\exp\left(i\lambda_{Pm}\right)}{2\Delta}\right]\exp\left[-\frac{\exp\left(-i\lambda_{Pj}\right)}{2\Delta}\right]\right)\nonumber \\
 & =\left(-1\right)^{N-1}\left(-2\Delta\right)^{\left(N-1\right)N/2}\prod_{k}\exp\left(i\left(N-k\right)\lambda_{Pk}\right)\prod_{j<m}\exp\left(-\frac{\cos\left(\lambda_{Pj}\right)+\cos\left(\lambda_{Pm}\right)}{2\Delta}\right)\prod_{j<l}\exp\left(-i\frac{\sin\left(\lambda_{Pj}\right)-\sin\left(\lambda_{Pm}\right)}{2\Delta}\right)\nonumber \\
 & =\left(-1\right)^{N-1}\left(-2\Delta\right)^{\left(N-1\right)N/2}\prod_{k}\exp\left(iN\lambda_{k}\right)\prod_{j}\exp\left(-\left(N-1\right)\frac{\cos\left(\lambda_{Pj}\right)}{2\Delta}\right)\prod_{j}\exp\left(-i\left(N-2j\right)\frac{\sin\left(\lambda_{Pj}\right)}{2\Delta}-i\left(j\lambda_{Pj}\right)\right)\label{eq:Simplifications}
\end{align}
 Then we have that:
\begin{align}
 & \frac{1}{\left(2\Delta\right)^{N\left(N-1\right)}}\left|\left\langle \lambda_{1},....\lambda_{N}\mid\Psi_{0}\right\rangle \right|^{2}\nonumber \\
 & =\left[\det\left(\left[\exp\left(-i\left(l-1\right)\lambda_{i}+i\frac{\sin\left(\lambda_{i}\right)}{\Delta}\right)\right]^{j}\right)\right]^{2}\prod_{j}\exp\left(-\left(N-1\right)\frac{\cos\left(\lambda_{j}\right)}{\Delta}\right)\nonumber \\
 & =\prod_{j>i}\left|\exp\left(-i\left[\left(l-1\right)\lambda_{i}-\frac{\sin\left(\lambda_{i}\right)}{\Delta}\right]\right)-\exp\left(-i\left[l\lambda_{j}-\frac{\sin\left(\lambda_{j}\right)}{\Delta}\right]\right)\right|^{2}\prod_{j}\exp\left(-\left(N-1\right)\frac{\cos\left(\lambda_{j}\right)}{\Delta}\right)\nonumber \\
 & =\exp\left(-\left(N-1\right)\sum_{j}\frac{\cos\left(\lambda_{j}\right)}{\Delta}\right)\prod_{j>i}4\sin^{2}\left(\frac{l-1}{2}\left(\lambda_{i}-\lambda_{j}\right)-\frac{\sin\left(\lambda_{i}\right)-\sin\left(\lambda_{j}\right)}{2\Delta}\right)\label{eq:Vandermonde_XXZ}
\end{align}
 Now we have that the Bethe equations are given by:
\begin{align}
\exp\left(i\lambda_{j}L\right) & =\left(-1\right)^{N-1}\prod_{j\neq k}\frac{\exp\left(i\left(\lambda_{j}+\lambda_{k}\right)\right)+1-2\Delta\exp\left(i\lambda_{j}\right)}{\exp\left(i\left(\lambda_{j}+\lambda_{k}\right)\right)+1-2\Delta\exp\left(i\lambda_{k}\right)}\nonumber \\
 & \cong\left(-1\right)^{N-1}\prod_{j\neq k}\exp\left(i\left[\lambda_{j}-\lambda_{k}\right]\right)\left(1-\frac{i}{\Delta}\left(\sin\left(\lambda_{k}\right)-\sin\left(\lambda_{j}\right)\right)\right)\nonumber \\
 & \cong\left(-1\right)^{N-1}\prod_{j\neq k}\exp\left(i\left[\lambda_{j}-\lambda_{k}\right]\right)\exp\left(-\frac{i}{\Delta}\left(\sin\left(\lambda_{k}\right)-\sin\left(\lambda_{j}\right)\right)\right)\nonumber \\
\Rightarrow\lambda_{j}\left(L-N\left(1+\frac{1}{\Delta}\frac{\sin\left(\lambda_{j}\right)}{\lambda_{j}}\right)\right) & =2\pi\left(I_{j}-\left(N-1\right)/2\right)-\sum_{k}\left[\lambda_{k}+\frac{1}{\Delta}\sin\left(\lambda_{k}\right)\right]\nonumber \\
\Rightarrow\left(l-1\right)\left(\lambda_{j}-\lambda_{i}\right) & =\frac{2\pi}{N}\left(I_{j}-I_{i}\right)+\frac{1}{\Delta}\left(\sin\left(\lambda_{j}\right)-\sin\left(\lambda_{i}\right)\right)\label{eq:Relationship_Bethe}
\end{align}
(with $I_{j}\in\mathbb{Z}$). This implies that
\begin{equation}
\frac{1}{\left(2\Delta\right)^{N\left(N-1\right)}}\left|\left\langle \lambda_{1},....\lambda_{N}\mid\Psi_{0}\right\rangle \right|^{2}=\exp\left(-\left(N-1\right)\sum_{j}\frac{\cos\left(\lambda_{j}\right)}{\Delta}\right)\prod_{j>i}4\sin^{2}\left(\frac{\pi}{N}\left(I_{i}-I_{j}\right)\right)^{2}\label{eq:Simlified}
\end{equation}
We again see that we must have that every $I=0,1,2,....N-1$ (mod
$N)$ is occupied exactly once or the overlap is zero and:
\begin{equation}
\frac{1}{\left(2\Delta\right)^{N\left(N-1\right)}}\left|\left\langle \lambda_{1},....\lambda_{N}\mid\Psi_{0}\right\rangle \right|^{2}=\exp\left(-\left(N-1\right)\sum_{j}\frac{\cos\left(\lambda_{j}\right)}{\Delta}\right)\prod_{j>i}4\sin^{2}\left(\frac{\pi}{N}\left(i-j\right)\right)+O\left(\frac{1}{\Delta^{2}}\right)\label{eq:Simplified}
\end{equation}
if 
\begin{equation}
\left(\lambda_{j}-\lambda_{i}\right)=\frac{2\pi}{L-N}\left(j-i\right)-\frac{N}{\Delta\left(L-N\right)}\left(\sin\left(\lambda_{i}\right)-\sin\left(\lambda_{j}\right)\right)+\frac{2\pi N}{L-N}\left[m_{j}-m_{i}\right],\:m_{k}\in\mathbb{Z}\label{eq:Same_before}
\end{equation}
and zero otherwise. Or we return to the same richness of overlaps
as the Mott insulator quench. Again the quench action fails as Eq.
(\ref{eq:Equals}) fails. The case of the N$\grave{\mathrm{e}}$el
state, $l=2$, requires special attention. In this case the terms
in Eq. (\ref{eq:Same_before}) $\frac{2\pi N}{L-N}\left[m_{j}-m_{i}\right]\in2\pi\mathbb{Z}$
so may be neglected. Then the $I_{i}$ are uniquely determined with
no freedom and we get that:
\begin{equation}
\lambda_{j}N-\frac{N}{\Delta}\sin\left(\lambda_{j}\right)=2\pi\left(j-\frac{N-1}{2}\right)-\sum_{k}\left[\lambda_{k}+\frac{1}{\Delta}\sin\left(\lambda_{k}\right)\right],\:j=0,...N-1\label{eq:Simplified-1}
\end{equation}
We now note that $\lambda_{j}=-\lambda_{N-1-j}$ the equations simplify
to
\begin{equation}
\lambda_{j}-\frac{1}{\Delta}\sin\left(\lambda_{j}\right)=\frac{2\pi}{N}\left(j-\frac{N-1}{2}\right)=-\left[\lambda_{N-1-j}-\frac{1}{\Delta}\sin\left(\lambda_{N-1-j}\right)\right]\label{eq:Paring}
\end{equation}
and we obtain the pairing structure in Refs. \citep{Wouters_2014,Wouters_2015}.
\end{widetext}

\end{document}